\documentclass[prc,aps,twocolumn,superscriptaddress]{revtex4-2}
\usepackage{graphicx}
\usepackage{natbib}
\usepackage{xcolor}

\usepackage{subfig}
\usepackage{float}
\usepackage{amsmath}
\usepackage{hyperref}
\usepackage{changes} 
\usepackage{appendix}
\usepackage{bm}
\usepackage{lipsum}

\begin{document}

\title{Low-energy $^{17}$O($n,\gamma$)$^{18}$O reaction within the microscopic potential model and its role for the weak r-process}

\author{Nguyen Le Anh}
\email{anhnl@hcmue.edu.vn}
\affiliation{Department of Physics, Ho Chi Minh City University of Education, 280 An Duong Vuong, Cho Quan Ward, Ho Chi Minh City, Vietnam}

\author{Jasmine Sarahi Andrews}
\affiliation{California State University East Bay,
25800 Carlos Bee Boulevard, Hayward, CA 94542 }
\affiliation{Lawrence Livermore National Laboratory, 7000 East Avenue, Livermore, CA 94550}

\author{Bui Minh Loc}
\email{lmbui@sdsu.edu}
\affiliation{ San Diego State University,
5500 Campanile Drive, San Diego, CA 92182}

\author{Andre Sieverding}
\email{Corresponding author: sieverding2@llnl.gov}
\affiliation{Lawrence Livermore National Laboratory, 7000 East Avenue, Livermore, CA 94550}

\begin{abstract}
The neutron radiative capture reaction $^{17}\mathrm{O}(n,\gamma)^{18}\mathrm{O}$ plays a pivotal role in both nuclear structure studies and astrophysical nucleosynthesis, particularly in the formation of elements during hydrostatic and explosive stellar environments. We calculated the $^{17}\mathrm{O}(n,\gamma)^{18}\mathrm{O}$ cross section within the Skyrme Hartree-Fock potential model and analyzed electric dipole $E1$ transitions to both positive- and negative-parity states below the alpha-decay threshold in $^{18}\mathrm{O}$. Our cross sections are significantly different from the data available in commonly used libraries. We further investigate the impact of the new calculated cross section on weak $r$-process nucleosynthesis using large-scale reaction network calculations across a wide range of electron fractions and entropies. Our results show that the $^{17}\mathrm{O}(n,\gamma)^{18}\mathrm{O}$ reaction rate significantly influences the production of first $r$-process peak elements, such as strontium, under specific astrophysical conditions. This study highlights the importance of accurate nuclear data for light isotopes in modeling heavy-element synthesis and provides updated reaction rates for future nucleosynthesis simulations.
\end{abstract}

\maketitle

\section{Introduction}
\begin{figure}
    \centering
    \includegraphics[width=0.95\linewidth]{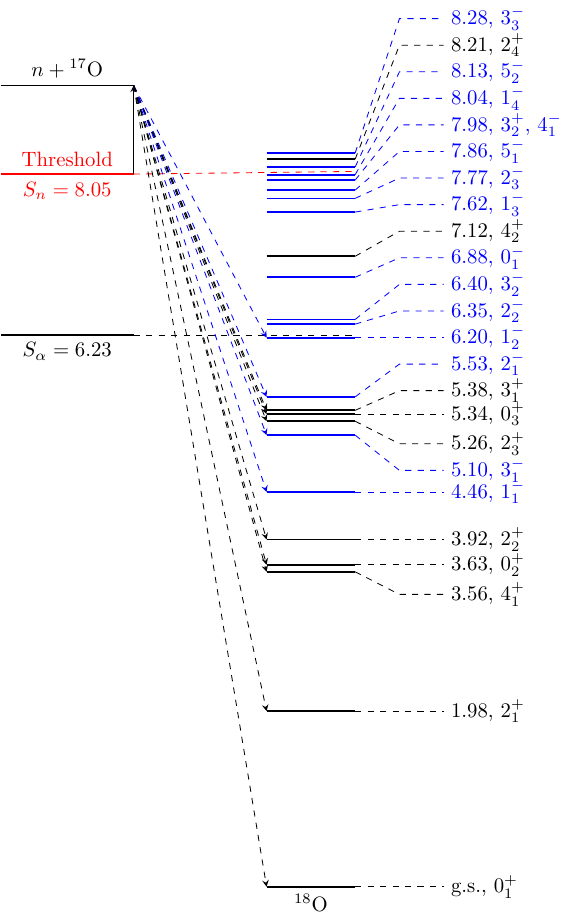}
    \caption{The structure of $^{18}\mathrm{O}$ and $E1$ transitions. The black dashed arrows indicate the transitions $J_s^- \to J^+_b$, while the blue dashed arrows indicate the transitions $J_s^+ \to J^-_b$. The energies are in MeV.}
    \label{fig:18O}
\end{figure}

In the reaction networks that describe the nucleosynthesis of the elements in the universe, all isotopes, from individual nucleons to the heaviest elements, are connected. Therefore, even reactions on light isotopes can have a large impact on heavy-element nucleosynthesis.
The $^{17}\mathrm{O}(n,\gamma)^{18}\mathrm{O}$ reaction is such a reaction that is known to be a crucial process in hydrostatic burning phases of stellar evolution, as pointed out by \citet{yamamoto2010} and that is also important for the nucleosynthesis of the elements heavier than iron in explosive astrophysical events. 
In core-collapse supernovae and neutron-star mergers, neutron-rich material is released and recombines, forming heavy elements. In this type of explosive nucleosynthesis, neutron captures on light isotopes, such as $^{17}\mathrm{O}$, are crucial as pathways to the formation of the initial seed nuclei, most prominently iron and nickel, on which neutrons are captured later in the evolution. As a consequence, reactions on light isotopes can have large implications on the final abundances of heavy elements. While the abundance pattern of the heaviest elements ($A>100$) shows a relatively robust pattern based on the observations of metal-poor stars \cite{Sneden.Cowan.ea:2008}, the abundances of the first $r$-process peak ($A\approx 80-90$) show larger variations in observations \cite{Montes.Beers.ea:2007}, making it more difficult to pinpoint the astrophysical site or site of the origin of the first $r$-process peak, often referred to as weak $r$-process. Furthermore, the identification of Sr ($Z=38$) in the aftermath of the kilonova associated with GW170817 \citep{Watson.ea:2019}
adds to the importance of the first $r$-process peak, including the cross sections for the nuclear reactions that shape it.
Beyond its astrophysical relevance, the $^{17}\mathrm{O}(n,\gamma)^{18}\mathrm{O}$ reaction is of considerable interest in nuclear structure studies. Studying this reaction at low neutron energies allows for understanding the resonance structure of $^{18}\mathrm{O}$, including the characterization of resonant states near the neutron separation energy (see Figure~\ref{fig:18O}).

Nuclear astrophysics studies require data on reaction cross sections across a wide range of isotopes.
For heavy nuclei with a large number of nuclear levels available around the neutron separation energy to capture into, the Hauser-Feshbach statistical model \citep{hauser1952} is applicable and there are several libraries of reaction rates available \citep{Rochman.Koning.ea:2025}. For light isotopes or weakly bound systems, however, reactions proceed via direct (nonresonant) capture into bound states and resonant capture into the few available low-lying states. 

For the $^{17}\mathrm{O}(n,\gamma)^{18}\mathrm{O}$ reaction, we employ a potential model which interprets the nucleon radiative capture process as electromagnetic transitions from scattering states to bound states, with the electric dipole ($E1$) transition being the most dominant \cite{huang2010}.
The potential model, though simple, is a powerful tool for deriving both ($n,\gamma$) and ($p,\gamma$) cross sections. The potential can be obtained in several ways. 
Ref.~\cite{zhang2022} used the RADCAP code \cite{bertulani2003} for direct capture cross section calculations, employing the phenomenological Woods-Saxon potentials for central and spin-orbit terms, along with a uniform Coulomb potential. However, phenomenological methods can introduce significant uncertainties due to the number of parameters involved, especially when the experimental data for the system is scarce.
The data for the calculation of the $^{17}\mathrm{O}(n,\gamma)^{18}\mathrm{O}$ cross section is surprisingly sparse, where only the thermal ($n,\gamma$) cross section has been directly measured as $\sigma_{\text{th}} = 0.67 \pm 0.07$ mb \citep{firestone2016} and there are two known resonances near the threshold at 8.21 MeV and 8.28 MeV \citep{Koehler.Graff:1991}. 
In contrast to the phenomenological approaches, we use a microscopic Skyrme Hartree-Fock potential \cite{anh2021, anh2021_2, anh2022, anh2022_2, anh2023}.

The paper is organized as follows: In Sec.~\ref{sec:method}, we describe the Skyrme Hartree-Fock potential model applied to nucleon radiative capture. In Sec.~\ref{sec:configs}, the configurations for the bound states of both positive-parity and negative-parity states in $^{18}\mathrm{O}$ are presented. Then, in Sec.~\ref{sec:partial-wave}, the $E1$ transitions based on the given bound configurations are analyzed. Finally, we include spectroscopic factors extracted from experiments to obtain the total cross section of $E1$ transitions. We compared the result with recent work in Ref.~\cite{zhang2022} and the implications for heavy-element nucleosynthesis in Section~\ref{sec:astro}. Conclusions are drawn in Sec.~\ref{sec:conclusion}.

\section{Skyrme Hartree-Fock potential approach} \label{sec:method}
First, we denote the initial state $| [I \otimes (\ell_s \otimes s)j_s]J_sM_s\rangle$, and the final state $| [I \otimes (\ell_b\otimes s)j_b]J_bM_b \rangle$. The system is considered as the $^{17}\mathrm{O}$ core, with one additional nucleon in a single-particle (s.p.) state. The core is assumed to have an internal spin $\bm{I}$ that remains unchanged. The ground state of $^{17}\mathrm{O}$ is $I^\pi = 5/2^+$. The total relative angular momentum of the nucleon-target system is represented by $\bm{j} = \bm{\ell} + \bm{s}$, where $\bm{\ell}$ is the relative orbital angular momentum and $s = 1/2$ for the nucleon. The channel spin of the initial system is then given by $\bm{J} = \bm{I} + \bm{j}$.

The radiative-capture cross section is given by
\begin{equation}\label{sigmaE}
    \sigma(E) = \dfrac{4}{3}\dfrac{1}{\hbar v} \left(\dfrac{4\pi}{3} k^3_\gamma \right) \dfrac{1}{(2s+1)(2I+1)} \sum_{\ell_s j_s J_s} |\mathcal{M}_{E 1}|^2,
\end{equation}
where $v$ is the relative velocity of the colliding particles, and $k_\gamma$ is the photon wave number defined by $\hbar c k_\gamma = E + Q$, with $Q$ being the $Q$ value of the reaction.

$\mathcal{M}_{E1}$ can be reduced to the calculation of the s.p. reduced matrix element
\begin{equation}\label{ME12}
    \mathcal{M}_{E1} = C_{e} S_{\text{F}}^{1/2}
    (-1)^{I+j_s+J_s+1}
    \hat{J_s}\hat{J_b}
    \left\{ \begin{matrix} 
        j_b & J_b & I \\ J_s & j_s & 1 
    \end{matrix} \right\}
    \mathcal{M}_{E1}^{(\rm s.p.)},
\end{equation}
where the value of effective charge is $C_e=-0.448e$ in the $n+^{17}$O system (with $e$ being the elementary charge), and $\hat{l} = \sqrt{2l + 1}$. The curly bracket represents the Wigner $6j$ coefficient. The spectroscopic factor, $S_{\text{F}}$, accounts for any missing configuration. In the potential model for the radiative-capture reaction, $S_{\text{F}}$ is adjusted to match experimental data. 

The s.p. reduced matrix element $\mathcal{M}_{E1}^{(\rm s.p.)}$ consists of two main components.
\begin{equation}\label{3com}
    \mathcal{M}_{E1}^{(\rm s.p.)}
    = \mathcal{A}_{E1} \mathcal{I}_{E1}.
\end{equation}
The geometrical coefficient $\mathcal{A}_{E1}$ is given by
\begin{align} \label{Elam}
    \mathcal{A}_{E1} =
    \sqrt{\dfrac{3}{4\pi}}
    \hat{j_b} \hat{\ell_b} \hat{\ell_s}
    \left( \begin{matrix} 
        \ell_b & 1 & \ell_s  \\ 0 & 0 & 0 
    \end{matrix} \right)
    \left\{ \begin{matrix} 
        \ell_b & j_b & 1/2 \\ j_s & \ell_s & 1 
    \end{matrix} \right\},
\end{align}
where the round bracket denotes the $3j$ coefficient.
The radial overlap integral $\mathcal{I}_{E1}$ is defined as
\begin{equation}\label{IE1}
    \mathcal{I}_{E1} = \int \phi_{n \ell_b j_b}(r) r \chi_{\ell_s j_s}(E,r) \,dr.
\end{equation}
$\chi_{\ell_s j_s}$ and $\phi_{n \ell_b j_b}$ are the s.p. scattering and bound wave functions, respectively. They are simultaneously obtained in the Skyrme Hartree-Fock potential model \cite{anh2021}.

The neutron-nucleus HF potential consists of the central nuclear and the spin-orbit terms
\begin{equation} \label{eq:Vq}
    V(E,r) = V_0(E,r) + V_{\text{s.o.}}(r).
\end{equation}
The central potential is
\begin{align} \label{eq:V0}
V_0(E,r) &= \dfrac{m^*(r)}{m'} \bigg\{U_0(r) + \dfrac{1}{2} \dfrac{d^2}{dr^2}\left(\dfrac{\hbar^2}{2m^*(r)} \right) \nonumber \\
&- \dfrac{m^*(r)}{2\hbar^2} \left[\dfrac{d}{dr}\left(\dfrac{\hbar^2}{2m^*(r)}\right)\right]^2\bigg\} + \left[1-\dfrac{m^*(r)}{m'}\right]E, 
\end{align}
where $m' = mA/(A-1)$, with $m$ and $A$ being the nucleon mass and target mass number, respectively. The spin-orbit potential is 
\begin{align} 
    V_{\text{s.o.}}(r) &= \dfrac{m^*(r)}{m'} U_{\text{s.o.}}(r)(\vec{\ell}\cdot\vec{s}), \label{eq:Vso}
\end{align}
where $U_0(r)$, $U_{\text{s.o.}}(r)$, and the effective masses $m^*(r)$ are obtained from the Skyrme HF equations~\cite{vautherin1972, colo2013}. The distinction between the two states enters only through the energy: $E<0$ yields the discrete bound state $\phi_{n\ell_b j_b}$, whereas $E>0$ corresponds to the continuous scattering state $\chi_{\ell_s j_s}$.

In our approach, we have two parameters, $\lambda_s$ and $\lambda_b$, to adjust the depths $V_0$ of the nuclear scattering and bound potentials, respectively. $\lambda_s$ is adjusted to reproduce exactly the location of the s.p. resonances. $\lambda_b$ is adjusted to replicate the s.p. separation energy $S_n$ for the transition to the ground state. Note that for the transitions to excited states, the excitation energy is added to $S_n$.

\section{Configurations for bound state} \label{sec:configs}
The nuclear structure of $^{18}\mathrm{O}$ as depicted in Figure~\ref{fig:18O} is particularly complex. 
In the potential model, the structure of $^{18}\mathrm{O}$ is a system of $n+^{17}$O. The spectroscopic factors were derived from investigations into the $^{17}\mathrm{O}(d, p)^{18}\mathrm{O}$ reaction, as discussed in Ref.~\cite{li1976}.

In the potential model, $^{18}\mathrm{O}$ is a valence neutron with the $^{17}\mathrm{O}$ core at its ground state ($5/2^+$). It should be noted that transitions are considered only from the scattering states above the neutron separation energy $S_n = 8.045$ MeV to the states below the alpha-decay threshold $S_\alpha = 6.23$ MeV (see Figure~\ref{fig:18O}). Above $S_\alpha$, the neutron+core component is expected to be small due to the dominance of $\alpha$ configurations.

Cluster-model calculations suggest that the negative-parity states of $^{18}\mathrm{O}$ are better described by an $^{15}$O core with three neutrons in the $sd$ shell, than by an $^{16}$O core with two neutrons in the $sd$ and the $fp$ shells \cite{buck1978}. In our s.p. model for low-lying states, however, we describe positive-parity states with $s,d$ states. Negative-parity states are described as those formed by $p$ states, as we expect captures into the $fp$ shell to lead to higher excitation energies. 

We selected the SLy4 and SkP interactions for this study. The scaling factor $\lambda_b$ is applied to the depth of the bound-state potential to adjust the s.p. bound-state energy equal to the neutron separation energy of the nuclear state. Tables~\ref{tab:lambda_b1_plus_state} and~\ref{tab:lambda_b2_plus_state} present the scaling factor $\lambda_b$ values for the SLy4 and SkP interactions used for bound states in various configurations within the $sd$ shell. 
The values of $\lambda_b$ provide insight into the distribution of s.p. states in the low-lying states of $^{18}\mathrm{O}$. For example, the SLy4 interaction yields $\lambda_b = 1$ for the s.p. $1d_{5/2}$ state in the $2^+_1$ state at 1.98 MeV, indicating a purely s.p. $1d_{5/2}$ in this state. In contrast, for the SkP interaction, the presence of the $1d_{5/2}$ orbital contributes to the $0_1^+$ and $2_1^+$ states. These observations suggest that the spectroscopic factor for the bound $1d_{5/2}$ state is significant, approaching unity, for the first low-lying states in the $^{18}\mathrm{O}$ structure.

\begin{table}[]
    \centering
    \setlength{\tabcolsep}{5pt}
    \caption{Scaling factors $\lambda_b$ obtained using SLy4 and SkP interactions for possible s.p. bound-state configurations of the first four positive-parity states in $^{18}\mathrm{O}$.}
    \label{tab:lambda_b1_plus_state}
    \begin{tabular}{ccccc} 
    \hline \hline
        \multicolumn{5}{c}{SLy4}\\ 
        \hline 
        $E_x$ & $0$ ($0_1^+$) & $1.98$ ($2_1^+$) & $3.56$ ($4_1^+$) & $3.63$ ($0_2^+$)  \\ \hline 
        $1d_{5/2}$ & $1.05$ & $1.00$ & $0.91$ & $0.90$  \\
        $2s_{1/2}$ & - & $1.10$ & - & -  \\
        $1d_{3/2}$ & - & $1.29$ & $1.23$ & -  \\ \hline \hline
        \multicolumn{5}{c}{SkP}\\ 
        \hline 
        $E_x$ & $0$ ($0_1^+$) & $1.98$ ($2_1^+$) & $3.56$ ($4_1^+$) & $3.63$ ($0_2^+$)  \\ \hline 
        $1d_{5/2}$ & $1.03$ & $0.96$ & $0.90$ & $0.89$ \\
        $2s_{1/2}$ & - & $1.02$ & - & -  \\
        $1d_{3/2}$ & - & $1.21$ & $1.14$ & - \\ \hline \hline
    \end{tabular}
\end{table}

\begin{table}[]
    \centering
    \setlength{\tabcolsep}{5pt}
    \caption{Same as Table~\ref{tab:lambda_b1_plus_state} but for subsequent positive-parity states in $^{18}\mathrm{O}$.}
    \label{tab:lambda_b2_plus_state}
    \begin{tabular}{ccccccccc} 
    \hline \hline
        \multicolumn{5}{c}{SLy4}\\ 
        \hline 
        $E_x$ & $3.92$ ($2_2^+$) & $5.26$ ($2_3^+$) & $5.34$ ($0_3^+$) & $5.38$ ($3_1^+$)  \\ \hline 
        $1d_{5/2}$ & $0.90$ & $0.84$ & $0.83$ & $0.83$ \\
        $2s_{1/2}$ & $1.01$ & $0.94$ & - & $0.93$ \\
        $1d_{3/2}$ & $1.21$ & $1.15$ & - & $1.15$ \\ \hline \hline
        \multicolumn{5}{c}{SkP}\\ 
        \hline 
        $E_x$ & $3.92$ ($2_2^+$) & $5.26$ ($2_3^+$) & $5.34$ ($0_3^+$) & $5.38$ ($3_1^+$)  \\ \hline 
        $1d_{5/2}$ & $0.88$ & $0.83$ & $0.82$ & $0.82$ \\
        $2s_{1/2}$ & $0.94$ & $0.88$ & - & $0.88$ \\
        $1d_{3/2}$ & $1.13$ & $1.07$ & - & $1.06$ \\ \hline \hline
    \end{tabular}
\end{table}

Table~\ref{tab:lambda_b_minus_state} shows that the $\lambda_b$ values are significantly decreased, reflecting the influence of the $p$ shell on the low-lying states of $^{18}\mathrm{O}$. Four negative-parity states are listed, each corresponding to configurations where the $p$ states couple with the $5/2^+$ core of $^{17}\mathrm{O}$. The spectroscopic factors for the $p$ states are notably smaller than those for the $s$ and $d$ states. 

\begin{table}[]
    \centering
    \setlength{\tabcolsep}{5pt}
    \caption{Same as Table~\ref{tab:lambda_b1_plus_state} but for negative-parity states in $^{18}\mathrm{O}$.}
    \label{tab:lambda_b_minus_state}
    \begin{tabular}{ccccccccc} 
    \hline \hline 
        \multicolumn{5}{c}{SLy4} \\ \hline 
        $E_x$ & $4.46$ ($1_1^-$) & $5.10$ ($3_1^-$) & $5.53$ ($2_1^-$) & $6.20$ ($1_2^-$)  \\ \hline 
        $1p_{3/2}$ & $0.41$ & $0.39$ & $0.37$ & $0.34$ \\
        $1p_{1/2}$ & - & $0.59$ & $0.56$ & - \\
        \hline \hline
        \multicolumn{5}{c}{SkP} \\ \hline 
        $E_x$ & $4.46$ ($1_1^-$) & $5.10$ ($3_1^-$) & $5.53$ ($2_1^-$) & $6.20$ ($1_2^-$)  \\ \hline 
        $1p_{3/2}$ & $0.51$ & $0.48$ & $0.46$ & $0.43$ \\
        $1p_{1/2}$ & - & $0.62$ & $0.60$ & - \\
        \hline \hline 
    \end{tabular}
\end{table}

Table~\ref{tab:SF} displays the spectroscopic factors extracted from Ref.~\cite{li1976}. The number of configurations has been significantly reduced compared to the total possible configurations. For the first three low-lying states, the spectroscopic factors for the $1d_{5/2}$ states are close to unity, indicating their dominant role. In contrast, the $2s_{1/2}$ states with large spectroscopic factors are predominantly located at the $3_1^+$ state. The contribution of $1d_{3/2}$ states to these positive and negative parity states is minimal. The negative-parity states primarily consist of $1p_{3/2}$ states, which exhibit very small spectroscopic factors. The $3^-_1$ state is composed of both $1p_{1/2}$ and $1f_{7/2}$ states, with the $1f_{7/2}$ state being more dominant. Neither \citep{li1976} nor the more recent compilation by \citet{tilley1995} contains data for the $2^-_1$ state at 5.53 MeV, so we assume the spectroscopic factor for the $1^-_2$ state at 6.20 MeV for the $2^-_1$ state. Despite their small values, the spectroscopic factors for the $p$ states are crucial in transitions from $s$ waves, which are key contributors to low-energy radiative capture processes.

\begin{table}[]
    \centering
    \setlength{\tabcolsep}{15pt}
    \caption{Spectroscopic factors extracted from Ref.~\cite{li1976}. The  $2_1^-$ state at 5.53 MeV is not included in the data.}
    \label{tab:SF}
    \begin{tabular}{cccc}
    \hline\hline
        $E_x$ & $J_b^\pi$ & $n\ell_b j_b$ & $S_{\rm F}$ \\ \hline
        $0$ & $0^+_1$ & $1d_{5/2}$ & $1.22$ \\ 
        $1.98$ & $2^+_1$ & $1d_{5/2}$ & $0.83$ \\
        & & $2s_{1/2}$ & $0.21$ \\ 
        $3.56$ & $4^+_1$ & $1d_{5/2}$ & $1.57$ \\ 
        $3.63$ & $0^+_2$ & $1d_{5/2}$ & $0.28$ \\
        $3.92$ & $2^+_2$ & $1d_{5/2}$ & $0.35$ \\
        & & $2s_{1/2}$ & $0.66$ \\ 
        $4.46$ & $1^-_1$ & $1p_{3/2}$ & $0.03$ \\
        $5.10$ & $3^-_1$ & $1p_{1/2}$ & $\leq 0.02$ \\
        & & $1f_{7/2}$ & $0.03$ \\
        $5.26$ & $2^+_3$ & $1d_{5/2}$ & $\sim 0$ \\
        & & $2s_{1/2}$ & $0.35$ \\ 
        $5.34$ & $0^+_3$ & $1d_{5/2}$ & $0.16$ \\ 
        $5.38$ & $3^+_1$ & $2s_{1/2}$ & $1.01$ \\
        & & $1d_{3/2}$ & $\sim 0$ \\ 
        $6.20$ & $1^-_2$ & $1p_{3/2}$ & $0.03$ \\ \hline \hline

    \end{tabular}
\end{table}

\section{Partial-wave analysis} \label{sec:partial-wave}

\begin{figure}
    \centering
    \includegraphics[width=0.95\linewidth]{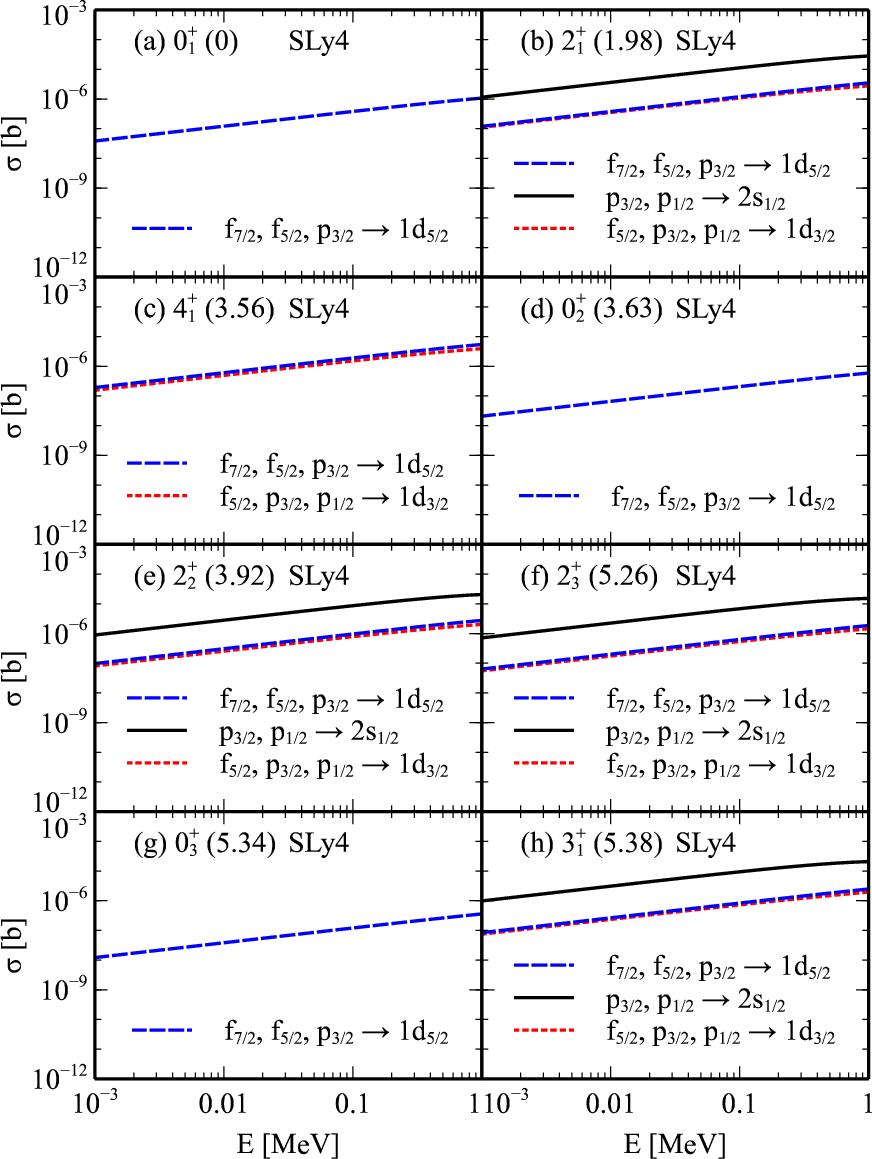}
    \caption{The partial-wave analysis for the $E1$ transitions from $J^-_s \to J^+_b$ in $^{17}\mathrm{O}(n,\gamma)^{18}\mathrm{O}$ with SLy4 interaction.}
    \label{fig:SLy4_plus}
\end{figure}

The scaling factors $\lambda_s$ for the scattering states are set to unity in all cases, indicating that the Skyrme Hartree-Fock mean field provides a reliable description of these continuum states~\cite{anh2021,anh2021_2,anh2022,anh2022_2,anh2023}. Figure~\ref{fig:SLy4_plus} shows the possible transitions to the ground state and various excited states of positive-parity states for the SLy4 interactions. The dominant transitions occur from scattering $p$ waves to the $2s_{1/2}$ state, while transitions from $p$ waves to the $1d_{5/2}$ state are roughly ten times less significant. This difference arises because the bound $s$ wave lacks a centrifugal barrier, making the $p \to 2s_{1/2}$ transition more favorable. Although transitions from $f$ waves to $d$ waves occur, they are much less significant compared to the $p \to 2s_{1/2}$ and $p \to 1d_{5/2}$ transitions and are not shown in the figure. The $E1$ transitions from $p$ waves to s.p. states in the $sd$ space exhibit little sensitivity to the choice of Skyrme interaction. In all these calculations, a spectroscopic factor of 1.00 is used. The value of the spectroscopic factor directly impacts the contribution of each transition.

\begin{figure}
    \centering
    \includegraphics[width=0.95\linewidth]{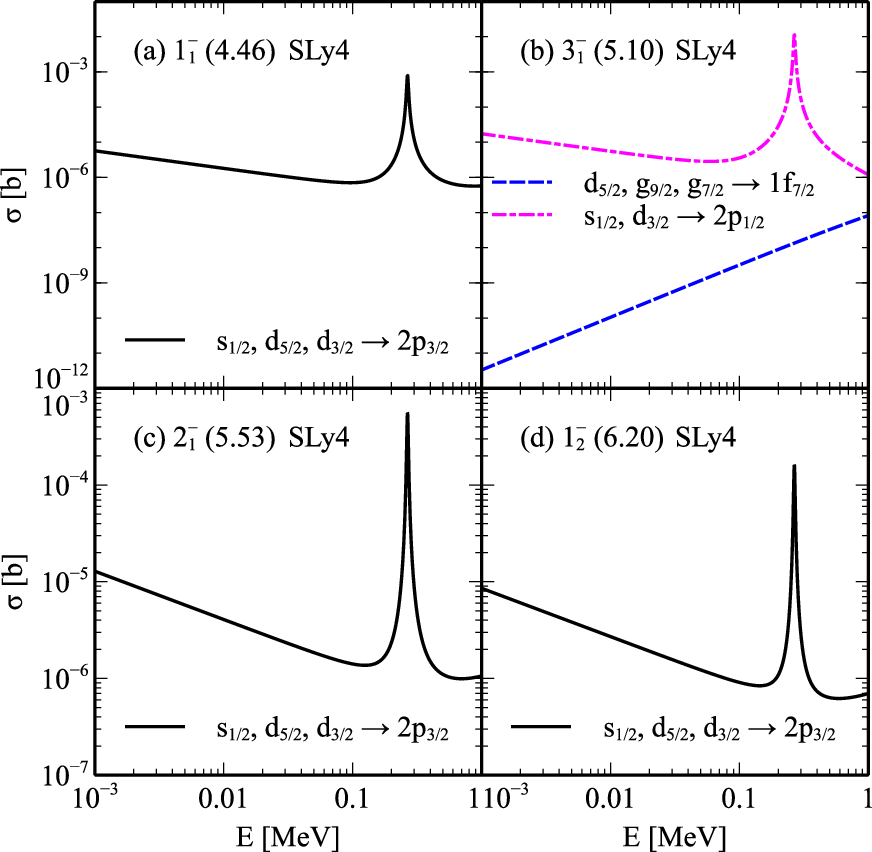}
    \caption{The partial-wave analysis for the $E1$ transitions from $J^+_s \to J^-_b$ in $^{17}\mathrm{O}(n,\gamma)^{18}\mathrm{O}$ with SLy4 interaction. A resonance is observed at low energy near the threshold, attributed to the scattering $d_{3/2}$ wave. }
    \label{fig:SLy4_minus}
\end{figure}

\begin{figure}
    \centering
    \includegraphics[width=0.95\linewidth]{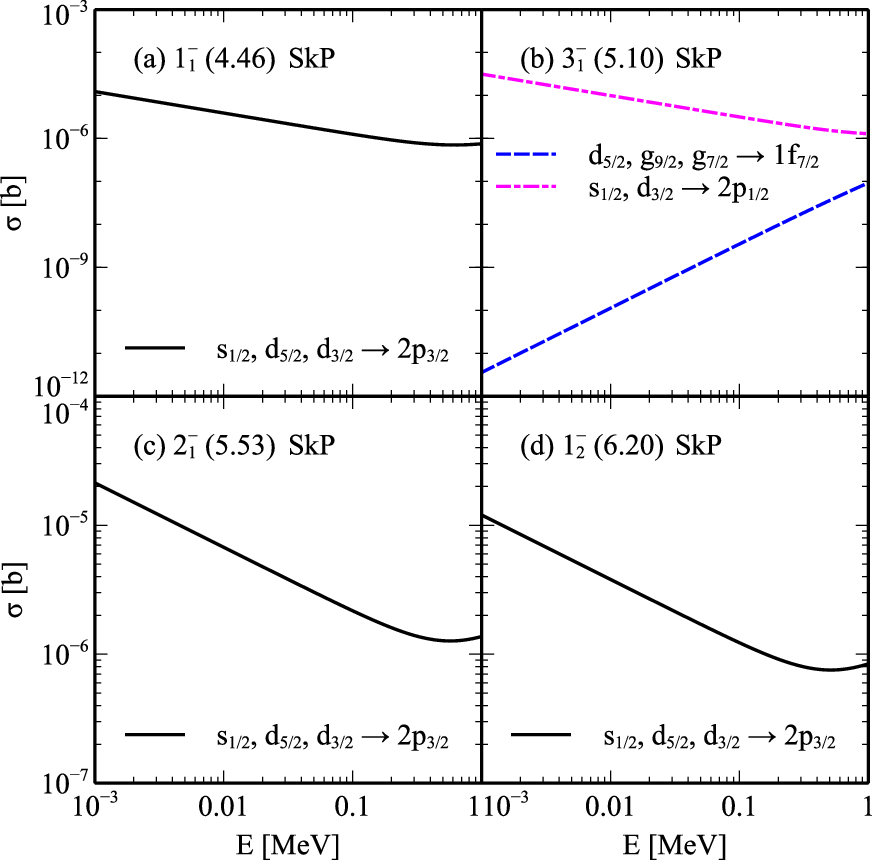}
    \caption{Same as Figure~\ref{fig:SLy4_minus} but for SkP interaction. No resonance is observed above the threshold, the resonance is located below the threshold.}
    \label{fig:SkP_minus}
\end{figure}

Figures~\ref{fig:SLy4_minus} and \ref{fig:SkP_minus} illustrate a significant difference between the SLy4 and SkP interactions. Specifically, the $d_{3/2}$ resonant scattering appears in the SLy4 interaction but is absent in the SkP interaction. In our calculation, this resonance is located at 240 keV, while experimentally, an excited  $2_4^+$ state is observed at just 160 keV above the threshold. Additionally, there is an ambiguous excited state around 7.98 MeV, possibly corresponding to either the $3_2^+$ or $4_1^-$ states. This resonance is very close to the threshold, leading some Skyrme interactions to produce a resonance pattern. 

It is important to note that the presence of the $d_{3/2}$ resonance has minimal impact on the cross section at low energies, which are of particular interest in astrophysics. Even transitions involving this resonance show a slight decrease in cross section at low energies. This can be attributed to the narrow width of the s.p. resonance. The primary contribution at low energies comes from transitions from the $s$ states to $p$ states.

Sub-threshold resonances in radiative capture play a crucial role at low energies, alongside resonances near and above the threshold. Their presence leads to an enhancement of the cross section at low energies, which, in turn, increases the nuclear reaction rate. The scattering $s_{1/2}$ resonance below the threshold makes the most significant contribution in transitions from $s$ states to $p$ states, compared to transitions from $d$ states to $p$ states. 

Figure~\ref{fig:18O} illustrates the states of $^{18}\mathrm{O}$ in the vicinity of the threshold. The state at 7.98 MeV is identified as either $3^+_2$ or $4^-_1$. If this state corresponds to $3^+_2$, then the $s_{1/2}$ state can indeed cause the resonance.

In a manner similar to the treatment of transitions to negative-parity states as shown in Figures~\ref{fig:SLy4_minus} and \ref{fig:SkP_minus}, we adjusted $\lambda_s = 0.76$ and $\lambda_s = 0.70$ to reproduce the sub-threshold resonance at $-0.068$ MeV below the threshold for the SLy4 and SkP interactions, respectively. 

\section{Total cross section} \label{sec:cross-section}

\begin{figure}
    \centering
    \includegraphics[width=0.95\linewidth]{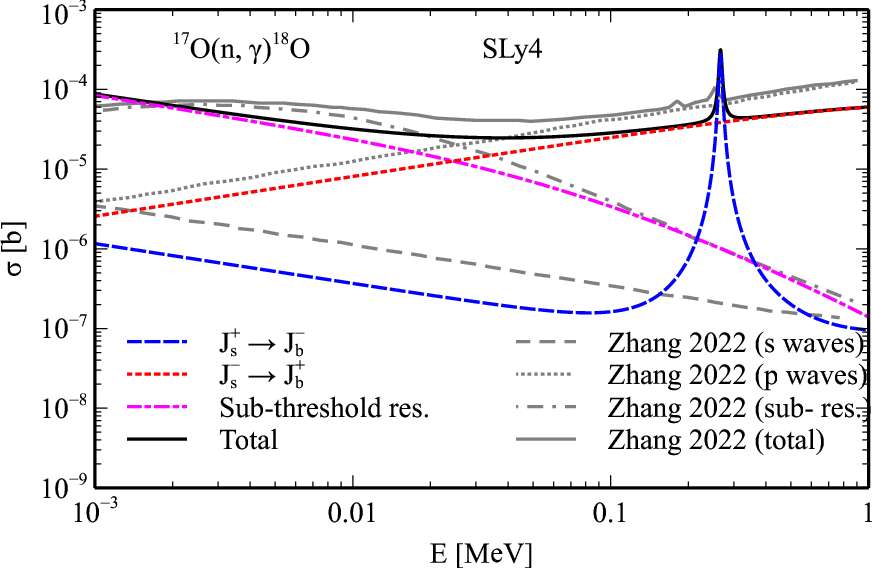}
    \caption{Cross section of the $^{17}\mathrm{O}(n,\gamma)^{18}\mathrm{O}$ reaction using the SLy4 interaction. The results are compared with those from Ref.~\cite{zhang2022}. The resonance is located near the resonances analyzed in Ref.~\cite{zhang2022}.}
    \label{fig:SLy4_all_SF}
\end{figure}

\begin{figure}
    \centering
    \includegraphics[width=0.95\linewidth]{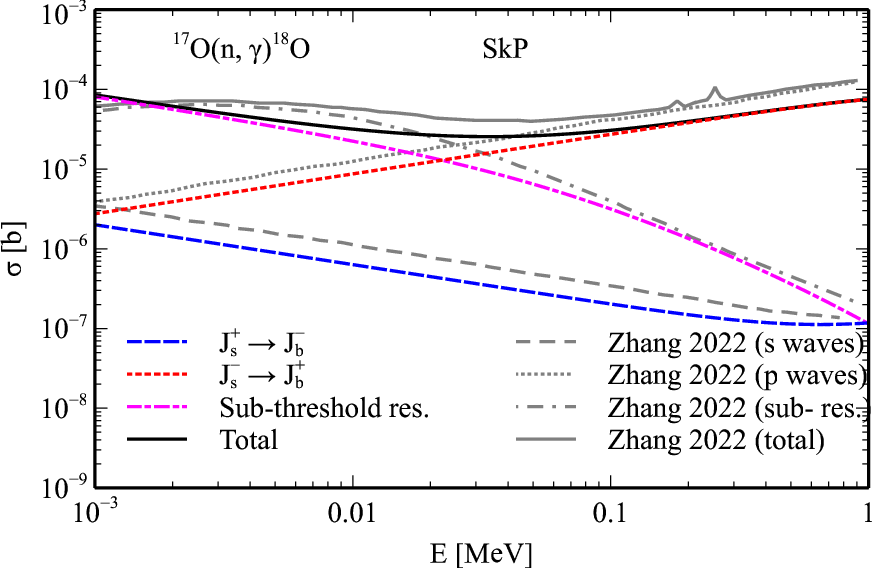}
    \caption{Same as Figure~\ref{fig:SLy4_all_SF} but for SkP interaction. The results show no resonance but are close to the direct capture calculated in Ref.~\cite{zhang2022}.}
    \label{fig:SkP_all_SF}
\end{figure}

Figures~\ref{fig:SLy4_all_SF} and \ref{fig:SkP_all_SF} show the calculated cross sections using the SLy4 and SkP interactions, respectively. The spectroscopic factors for the bound configurations, listed in Table~\ref{tab:SF}, are taken from Ref.~\cite{li1976}. Our calculations are compared with the results from Ref.~\cite{zhang2022}, which include three main contributions: direct capture from $s$ and $p$ waves, resonant capture, and sub-threshold resonance. Among these, the sub-threshold resonance contribution is the most significant.

It is noteworthy that our calculations predict the resonance in the SLy4 interaction case without requiring any modifications or additional calculations as shown in Figure~\ref{fig:SLy4_all_SF}. Furthermore, the transitions involving even-parity waves in our SLy4 calculations are lower than the $s$ wave transitions reported in Ref.~\cite{zhang2022}, whereas, with the SkP interaction, the transitions of even waves in our calculations are slightly higher. This difference is due to the presence of the $d_{3/2}$ resonance just below the threshold in the SkP case, which enhances the low-energy cross section.

The results for even and odd waves in our calculations show contrasting behaviors. Around 1.5 keV, their contributions are similar. Below 1.5 keV, even waves dominate, while odd waves become more dominant at higher energies. The sub-threshold resonance and direct capture from odd waves contribute equally to the cross sections around 35 keV. Below this energy, the cross section is strongly influenced by the sub-threshold resonance contribution.

Since the total cross sections are very similar on average for both interactions we have used, we consider only the cross section based on the SLy4 interaction for our astrophysical calculations.
Figure~\ref{fig:astro_rates} compares the astrophysical reaction rates we consider in this work. The astrophysical reaction rates are the reaction cross section folded with Maxwell-Boltzmann distributions for the relative motion of target and incident neutrons \citep{Pritychenko:2012,Pritychenko.ea:2010}. For this study, we do not take into account that the target can be in an excited state, an effect referred to as the stellar enhancement factor. This is justified for the case of $^{17}\mathrm{O}$ because the first excited state is located at 870~keV. 

Figure~\ref{fig:astro_rates} shows again that our results are close to \citet{zhang2022} and similarly the reaction rate is significantly higher than the reaction rates from REACLIB and lower than ENDF/B VIII.1. Note that the reaction rate in REACLIB is based on \citet{Koehler.Graff:1991}, who used an extrapolation based on the thermal neutron capture cross section and a measurement of the  $^{17}\mathrm{O}(n,\alpha)$ cross section. 
For comparison, we also consider the reaction rate from TENDL-2021 \citep{TENDL_2019}, which is a reaction cross section based on the Hauser-Feshbach statistical model code TALYS. For light isotopes such as $^{17}\mathrm{O}$, where typically few transitions dominate the reaction cross section, the statistical model is not applicable, and calculations such as the direct capture calculation presented here are more appropriate. Similar to the REACLIB reaction rate, TENDL-2021 cross section results in an astrophysical reaction rate that is significantly lower than the direct capture results.

\section{Role in weak \textit{r}-process nucleosynthesis}
\label{sec:astro}
The role of the reaction $^{17}$O($n,\gamma$)$^{18}$O for the origin of heavy elements has been discussed by \citet{zhang2022} in the context of weak and main $s$-process, as well as for the $r$-process in a collapsar model and a supernova-associated neutron burst. Since our new cross section is not significantly different from the findings in \citet{zhang2022}, we expect their conclusions to hold. Here, we investigate the $\alpha$ process and weak $r$ process that had not been considered by \citet{zhang2022}. The weak $r$-process occurs for moderately neutron-rich conditions between a pure $\alpha$ rich-freeze out and the ``strong'' $r$-process that is responsible for the heaviest elements. Such moderately neutron-rich conditions contribute to the first $r$-process abundance peak around $A \approx 90-100$. The conditions we consider are a subset of the conditions recently studied by \citep{Kuske.Arcones.ea:2025} with a similar parametric setup. The conditions could be found in various contexts, including neutron star mergers \citep{Sprouse.Lund.ea:2024,Just.Xiong.ea:2023} and core-collapse supernovae \citep{Wang.Burrows:2024,Glas.Just.ea:2018}. Therefore, it is particularly important to understand the detailed nucleosynthesis signatures that may serve to disentangle contributions from different sites. 

\subsection{Simulation Setup}
\begin{figure}
    \centering
    \includegraphics[width=\linewidth]{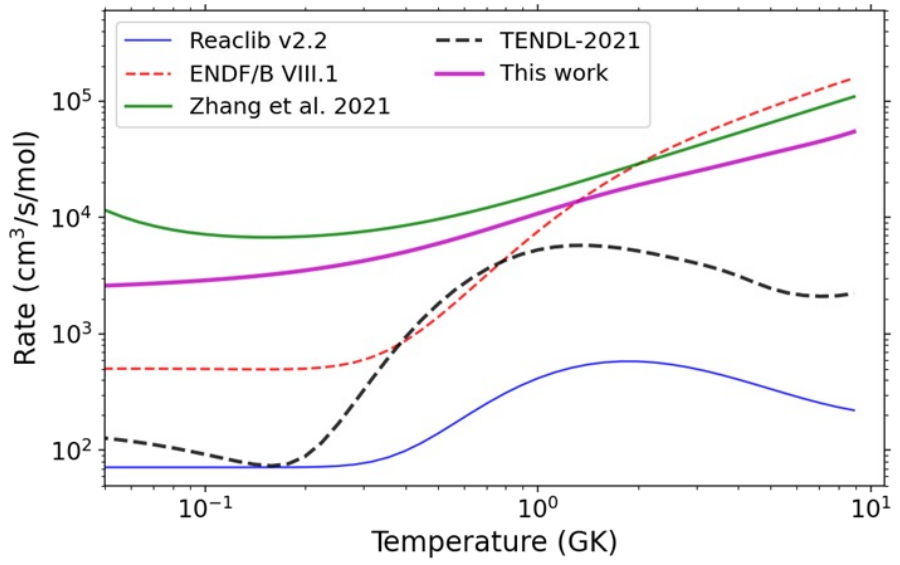}
    \caption{Astrophysical reaction rates based on cross sections from different sources compared to the calculations presented above.}
    \label{fig:astro_rates}
\end{figure}

We initialized calculations at Nuclear statistical equilibrium (NSE), treating initial electron fraction ($Y_e$) and initial specific entropy ($S_0$) as free parameters to investigate. We set the initial temperature at $T_0 = 10$ GK and keep a constant expansion time scale for this study $\tau = 10$~ms. We use the equation of state of \citep{Timmes.Swesty:2000} to determine the corresponding initial density $\rho_0$. We parameterize the density evolution of the ejecta as in \cite{Lippuner.Roberts:2015}:
\begin{equation}\rho(t) =\begin{cases}
    \rho_0 \, e^{-t/\tau} & \text{if $t\leq 3\tau$}\\[12pt]
    \rho_0 \, \left(\frac{3\tau}{et}\right)^3 & \text{if $t\geq 3\tau$},
\end{cases}
\end{equation}
where $e$ is Euler's number. 
To cover a wide range of possible scenarios, we explore a grid of initial  $ 0.2 \leq Y_e \leq 0.5 $ in steps of 0.01 and $10\, k_B/\mathrm{baryon}\leq S_0 \leq 100\, k_B/\rm{baryon}$ in steps of $1\,k_B/\mathrm{baryon}$.
On this grid of $Y_e$ and $S_0$, we perform nucleosynthesis calculations with reaction network code {\sc{XNet}} \footnote{https://github.com/starkiller-astro/XNet} that follows the evolution of the abundances $Y_i(t)$ of more than 7,000 isotopes, including all relevant reactions and decay channels and electron screening effects. 
Except for $^{17}$O$(n,\gamma)$ and its inverse reaction, all other reaction rates are taken from the Reaclib library \cite{Cyburt.Amthor.ea:2010}. We focus on the moderately neutron-rich, weak $r$-process regimes, so we do not include fission in our calculations.
We repeat the calculations with the reaction rates for $^{17}$O$(n,\gamma)$ as shown in Figure \ref{fig:astro_rates}. 
We do not include the effect of nuclear heating in the calculations to ensure that the thermodynamic variables are exactly the same for calculations with different nuclear reaction rates. 

\subsection{Overview of Conditions}

\begin{figure}
    \centering
    \includegraphics[width=\linewidth]{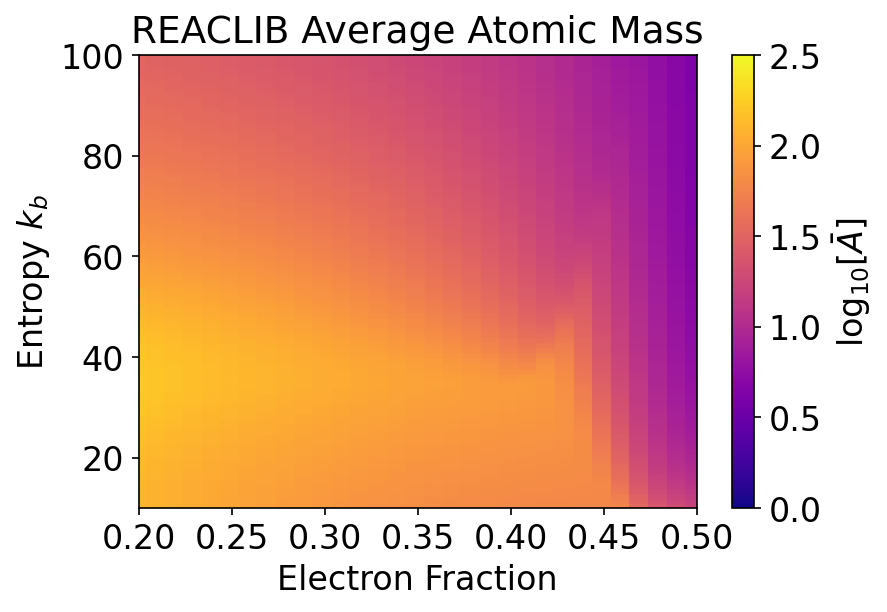}
    \caption{Average atomic mass number for the final abundances of the conditions considered in our study.}
    \label{fig:abar}
\end{figure}
Figure \ref{fig:abar} shows the mean atomic mass number across the range of astrophysical conditions, where 
$$ \bar{A} =\frac{ \Sigma_i A_i Y_i}{\Sigma A_i}.$$
Here, $i$ runs over all nuclear species in the reaction network. 
Initial values of $Y_e$ close to 0.5 are dominated by Fe-group elements from the freeze-out from NSE with low $\bar{A}$. Increasing entropy results in lower density for a given temperature, which slows down the conversion of $\alpha$ particle into $^{12}$C and leads to more leftover $^{4}$He at the end of the calculations. The large amount of leftover $\alpha$ particles decreases $\bar{A}$ with increasing entropy as seen in Figure \ref{fig:abar}.
At $Y_e\approx 0.46-0.48$, the heavy element nucleosynthesis up to the first $r$-process peak and including the neutron-deficient $p$-isotopes proceeds via the $\alpha$ process \citep{Hoffman.Woosley.ea:1996}, in which heavy elements are synthesized mainly through sequences of $\alpha$-particle induced reactions on neutron-rich seed nuclei that freeze out from NSE. This regime smoothly transitions into the weak $r$-process regime, where initial sequences of $(\alpha,n)$ reactions are increasingly followed by sequences of neutron captures that allow more and more heavy elements to form. Especially in the transition regime, $(\alpha,n)$ reactions play an important role \citep{Bliss.Arcones.ea:2017}.  
At low entropy and low $Y_e$, $\bar{A}$ reaches the highest values, indicating the conditions most suitable for the main $r$ process producing isotopes up to $A=200$. In the intermediate regime with $Y_e> 0.35$ and $S>30$ the first $r$-process peak is efficiently produced. 
The evolution of $\bar{A}$ as a function of the electron fraction, corresponding to a horizontal line in Figure \ref{fig:abar}, is not always monotonously increasing towards lower values of $Y_e$ without setup. Instead, Figure \ref{fig:abar} shows that, for entropies between 40 and 60, there is a narrow region of reduced $\bar{A}$ for $Y_e$ between $0.40$ and $0.45$. This region marks the transition between the efficient conversion of $\alpha$ particles to heavy nuclei via the $\alpha$~process on the right, higher $Y_e$ side to a regime dominated by neutron captures, which initially slows down the consumption of $\alpha$ particles. As we will show below, the $^{17}$O$(n,\gamma)$ reaction has an impact on this transition. 

\subsection{\texorpdfstring{The Role of $^{17}$O$(n,\gamma)$}{The Role of 17O(n,g)}}

\begin{figure}
    \centering
    \includegraphics[width=0.99\linewidth]{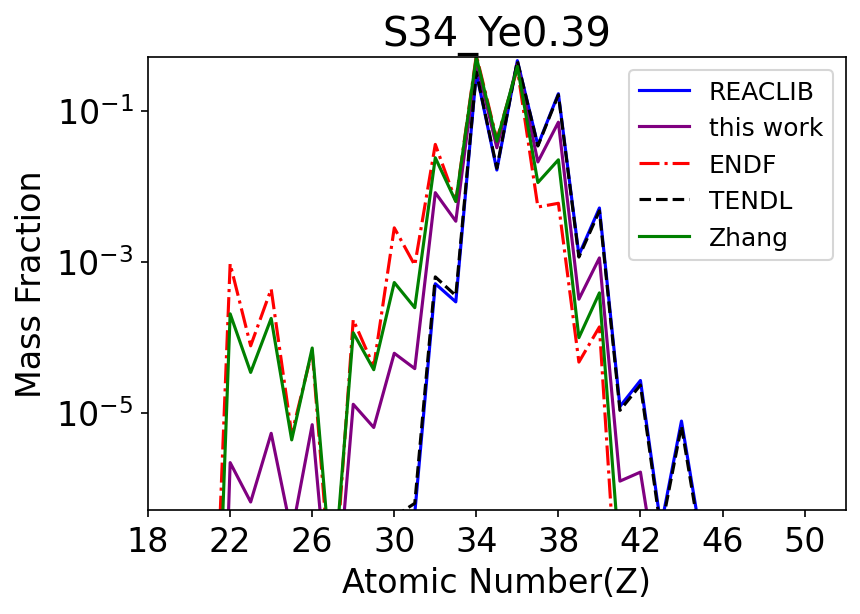}
    \caption{Abundance pattern for $Y_e=0.39$ and entropy $S=34k_{\rm{B}}/\rm{baryon}$
    where we find the largest impact of the $^{17}$O$(n,\gamma)$ reaction rate. 
    The abundance pattern around the first $r$-process peak
    is significantly affected by the reaction.}
    \label{fig:abundance_pattern}
\end{figure}

For specific combinations of $Y_e$ and entropy, we find a significant impact of the $^{17}$O$(n,\gamma)$ reaction rate on the final abundance pattern. 
For example, Figure \ref{fig:abundance_pattern} shows the final mass fractions for a calculation with $Y_e=0.39$ and $S=34\,k_{\rm{B}}$. 
The final composition shows a distinct peak for Se ($Z=34$) and Kr ($Z=36$). With the $^{17}$O$(n,\gamma)$ reaction rate from REACLIB, Sr ($Z=38$) is produced at a comparable level, and Kr is produced with a mass fraction larger than $10^{-3}$.  When the $^{17}$O$(n,\gamma)$ reaction rate is changed, the final mass fractions for Sr are significantly reduced. 
The difference in the results using the reaction rate from ENDF/B-VIII.1 and REACLIB is the largest. 
Using the $^{17}$O$(n,\gamma)$ reaction rate that results from the cross section in ENDF/B-VIII.1 leads to a Sr mass fraction more than a factor $10$ smaller than the result with REACLIB. Using the reaction rate by \citet{zhang2022} and our new result leads to final mass fractions for Sr in between, where our results are closer to REACLIB. 
This is consistent with the reaction rates shown in Figure \ref{fig:astro_rates}, where the difference between REACLIB and ENDF/B-VIII.1 is largest. 
From Figure \ref{fig:abundance_pattern} it is clear that the production of Sr and Kr is at the expense of the lighter elements. 
The final mass fractions for Fe ($Z=26$), Cr ($Z=24$) and Ti ($Z=22$) are below $10^{-5}$ with the reaction rate from REACLIB. On the other hand Ti reaches a mass fraction of $10^{-3}$ with the reaction rate from ENDF/B-VIII.1. 
These large differences in the production of the elements from Fe to Sr shown in Figure \ref{fig:abundance_pattern} are entirely the result of differences in the rate of $^{17}$O$(n,\gamma)$ and its inverse reaction. 

To better understand under which conditions the $^{17}$O$(n,\gamma)$ reaction has such a large impact on the production of the elements between $Z=22$ and $Z=42$, we calculate a root-mean-square deviation  between calculations with different reaction rates for each set of astrophysical parameters as 
\begin{equation}
    RMSD = \frac{1}{N}\sqrt{\sum_i^N \left(\frac{Y_i^a - Y_i^b}{Y_i^a}\right)^2},
    \label{eq:RMSD}
\end{equation}
where the superscripts $a$ and $b$ indicate results with different reaction rates and the index $i$ runs over all isotopes in the reaction network, excluding isotopes with abundances $ Y_i< 10^{-5}$. We have checked that the cutoff does not affect our conclusions, but setting a smaller cutoff leads to spurious contributions from isotopes with very small abundances.

\begin{figure}
    \centering
    \includegraphics[width=0.95\linewidth]{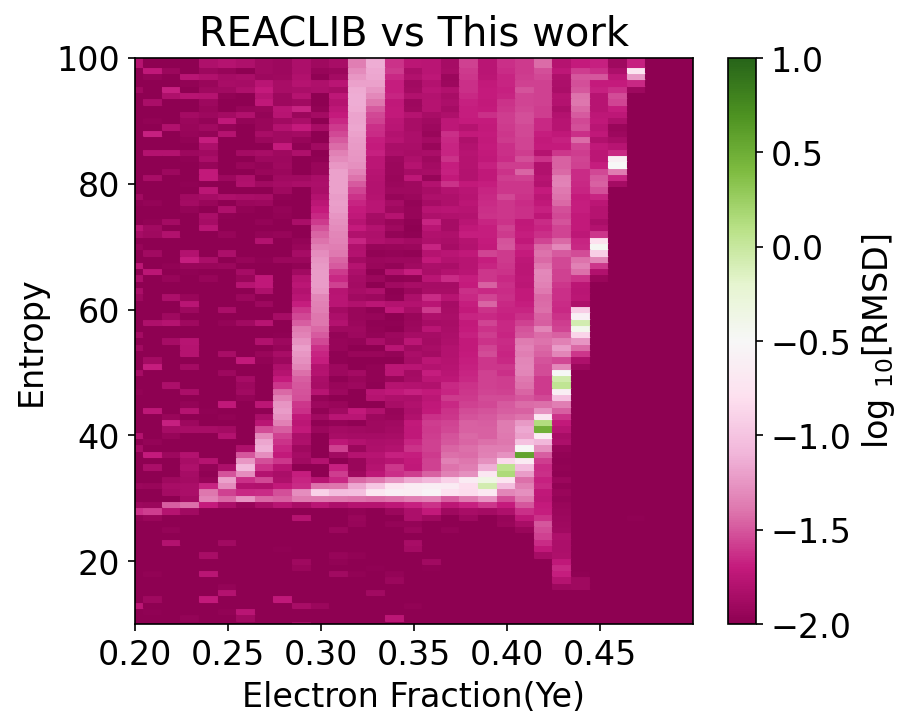}
    \caption{Root-mean-square difference for calculation with different $^{17}$O$(n,\gamma)$ reaction rate from REACLIB compared to calculations with our new reaction rate across the range of conditions for the weak $r$-process.}
    \label{fig:nucleo-heat-map}
\end{figure}

Figure \ref{fig:nucleo-heat-map} shows $RMSD$ from equation \ref{eq:RMSD} evaluated on our grid of simulations comparing the reaction rate from this work with the SLy4 interaction to simulation results with the REACLIB reaction rate. The overview  shows that there is significant structure in the differences, and we can clearly identify the narrow range of conditions where the 
$^{17}$O$(n,\gamma)$ reaction rate leads to particularly large $RMSD$. 
We find a nearly continuous line of large $RMSD$ that includes the set of conditions $S=34$, $Y=0.39$ that is shown in Figure \ref{fig:abundance_pattern}. The line of large $RMSD$ starts at nearly constant $S_0=25$ for low $Y_e$ up to about $Y_e=0.38$ and then curves sharply upward to reach $S_0=100$ at around $Y_e=0.47$.

The conditions where we find the largest impact are relevant for several astrophysical scenarios. 
For example, \citet{Wang.Burrows:2024} find similar conditions in low-mass core-collapse supernova explosions when modeled in 3D. In 3D simulations, the authors find that values of the specific entropy of up to $140 {k}_B$ can be reached, and electron fractions can be lower than $0.4$. Similarly, recent long-term simulations of neutron star mergers that include the effects of neutrinos indicate that both disk ejecta and dynamical ejecta are likely to contain material experiencing conditions with similar parameters \citep{Sprouse.Lund.ea:2024,Just.Xiong.ea:2023}. For example, \citet{Metzger.Fernandez:2014} found distributions of $Y_e$ centered around 0.4 in polar disk-wind outflows in simulations with long-lived hyper-massive neutron stars. However, such simulations come with large uncertainties, not only due to the implementation and choice of parameters, but also because the initial conditions are poorly constrained by observations. Thus, in order to accurately infer astrophysical conditions from nucleosynthesis observables, it is important to minimize uncertainties from the nuclear inputs.
 Figure \ref{fig:nucleo-heat-map} also shows a secondary line-like structure with lower $Y_e$ extending to high entropy at a much smaller scale of differences. 

The conditions with the largest $RMSD$ are characterized by transitions between reaction flow regimes in which the $^{17}$O$(n,\gamma)^{18}$O reaction becomes a major bottleneck in the reaction flow from $^{12}$C to Fe-group seed nuclei. The details of the changes in the reaction flow and what defines the conditions where $^{17}$O$(n,\gamma)^{18}$O is important are investigated below. 

\subsection{Impact on the Reaction Flow}

The conditions we are considering are mostly characterized by an $\alpha$-rich freeze-out from nuclear statistical equilibrium (NSE). An $\alpha$-rich freeze-out occurs when the cooling or expansion of matter is fast and not all free $\alpha$ particles can combine into $^{12}$C and heavier elements before the temperature and density are too low for the recombination to be complete. The result is a distribution of isotopes centered around Fe and Ni and a reservoir of free neutrons and $\alpha$ particles. 
Once reaction timescales are too short to establish an equilibrium state, the free particles are captured by the heavier ``seed'' nuclei, leading to the distinct nucleosynthesis processes, including the $r$ process. Once the $3\alpha$ reaction falls out of equilibrium, the remaining abundance of C and O tends to be converted to Fe and Ni by sequences of $(\alpha,n)$, $(\alpha,\gamma),$, and $(n,\gamma)$ reactions. The speed at which the conversion from light isotopes to Fe group isotopes occurs is also a non-equilibrium process, and it determines the availability of seed and free particle abundances when the $r$ and $\alpha$ processes operate. 

\begin{figure}
    \centering
    \includegraphics[width=0.99\linewidth]{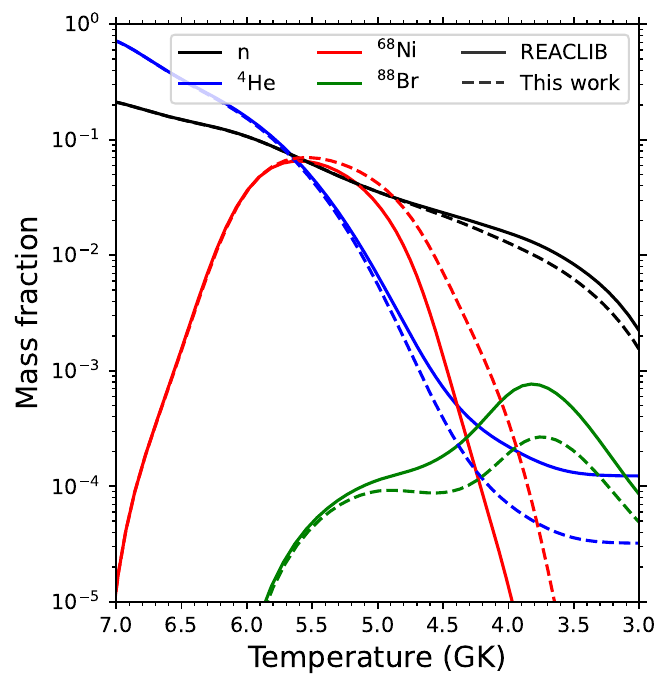}
    \caption{Evolution of mass fractions of neutrons, $\alpha$ particles, $^{68}$Ni and $^{88}$Br, which is the parent or $^{88}$Sr, for $Y_e=0.39$ and entropy $S=34 \,k_{\rm{B}}/\rm{baryon}$ for the calculation with the REACLIB $^{17}$O$(n,\gamma)^{18}$O and with our new reaction rate.  }
    \label{fig:evol_example}
\end{figure}

In the moderately neutron-rich conditions that we consider here, the reaction $^{17}$O$(n,\gamma)^{18}$O is one of the bottleneck reactions that regulate the conversion of carbon and $\alpha$ particles into Fe-group isotopes in the post-freeze-out phase. When $^{17}$O$(n,\gamma)^{18}$O is followed by $^{18}$O$(\alpha,n)^{21}$Ne an efficient pathway for further $\alpha$-induced reactions to the Fe-group exists. When  $^{17}$O$(n,\gamma)^{18}$O is reduced, on the other hand, reactions need to proceed through $^{17}$O$(\alpha,n)^{20}$Ne. Further  $\alpha$-induced reactions on $^{20}$Ne, however are slower than the reactions on $^{21}$Ne, reducing the consumption of $\alpha$ particles in the post-freeze-out phase.
This effect is illustrated in Figure \ref{fig:evol_example}, which shows the abundance evolution as a function of the decreasing temperature for  $^{68}$Ni and $^{88}$Br together with the free particle abundances for a calculation with our new $^{17}$O$(n,\gamma)^{18}$O reaction rate compared to the results using the reaction rate from REACLIB. $^{88}$Br eventually decays to $^{88}$Sr, and Figure \ref{fig:evol_example} shows the reduced final mass fraction of  $^{88}$Br for our new reaction rate, which results in a lower final abundance of Sr shown in Figure \ref{fig:abundance_pattern}. The radioactive isotope $^{68}$Ni is an intermediate product that is quickly produced from NSE seed nuclei, and that is either further transformed to heavier elements, or that decays to stable $^{68}$Zn. 
In contrast to $^{88}$Br, the mass fraction of  $^{68}$Ni is increased by our new reaction rate, and the $\alpha$ particle and neutron abundance are lower. The new reaction rate that is higher than the REACLIB value, thus favors the ongoing flow form C and O to Fe, consuming the available free particles earlier in a temperature range between 5 and 4 GK. The higher reaction rate from REACLIB reduces the flow out of O and leaves more free particles to convert Ni isotopes to heavier elements, including the radioactive parents of the stable Sr isotopes.  

Next, we investigate why the impact of the $^{17}$O$(n,\gamma)^{18}$O reaction rate is limited to a narrow range of conditions. 

The conditions under which $^{17}$O$(n,\gamma)^{18}$O is relevant are determined by the competition with alternative reaction pathways.
Instead of going to $^{18}$O, the reaction path can go directly from $^{17}$O to $^{20}$Ne, as mentioned above. When very few neutrons are available, the  $^{17}$O$(n,\gamma)^{18}$O reaction cannot happen, regardless of its cross section. In the other extreme case,  $^{18}$O can be directly produced from $^{15}$C, which requires more neutron-rich conditions to produce  $^{15}$C in the first place. 

\begin{figure}
    \centering
    \includegraphics[width=\linewidth]{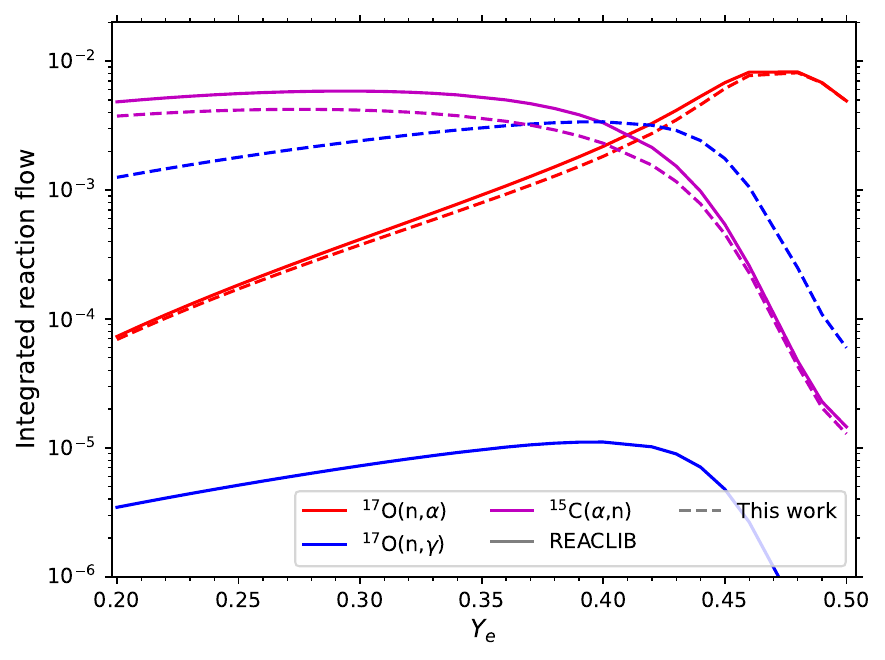}
    \caption{Reaction flows for $^{17}\mathrm{O}(n,\gamma)$  $^{17}\mathrm{O}(\alpha,n)$ and $^{15}\mathrm{C}(\alpha,n)$ integrated over the temperature range between 8 GK and 3 GK for our range of values for $Y_e$ and fixed $S=34 \,k_{\rm{B}}$/baryon. The capture reaction dominates the reaction flows only within a narrow range of $Y_e$ and only for our higher reaction rate.}
    \label{fig:ng_vs_na}
\end{figure}

Figure \ref{fig:ng_vs_na} illustrates the competition between the different reaction pathways depending on the initial conditions by showing the integrated net reaction flows across a range of $Y_e$ for a fixed value of the entropy $S_0 = 34 \,k_B$ with the $^{17}\mathrm{O}(n,\gamma)$ reaction rate from REACLIB (solid line) and with our new reaction rate. 
With the new reaction rate, the integrated flows show that the path through $^{17}\mathrm{O}(n,\gamma)$ is the dominant flow for values of $Y_e$ between 0.36 and 0.43, and this is the narrow range where the reaction rate can have a large impact on the final abundances. For values of $Y_e<0.36$,   $^{15}\mathrm{C}(\alpha,n)^{18}$O dominates, which means that the alternative pathways to $^{18}$O via more neutron-rich carbon isotopes are preferred. 
The alternative pathway includes a ``loop'' pathway through $^{17}$O$(n,\alpha)^{14}$C.
For $Y_e<0.36$, therefore, the $^{17}$O$(n,\gamma)$ reaction rate does not change the abundance pattern. For $Y_e>0.43$, on the other hand, the main reaction flow goes through $^{17}\mathrm{O}(\alpha,n)^{20}$Ne, avoiding the more neutron-rich isotopes of C and O altogether.
With the $^{17}\mathrm{O}(n,\gamma)$ rate from REACLIB, or ENDF/B-VIII.1 the path through $^{17}\mathrm{O}(n,\gamma)$ is much smaller than the other reaction paths for all values of $Y_e$ in our grid. 

The large impact of the  $^{17}\mathrm{O}(n,\gamma)$ reaction we see, is therefore due to a transition from the $\alpha$ particle dominate flow directly through $^{17}\mathrm{O}$ and the neutron capture dominated reaction path through $^{14,15}\rm{C}$.

The difference in the low $Y_e$ region around $Y_e=0.23$ and $S_0=30$ results from a similar effect. Towards more neutron-rich conditions, the bottleneck reaction shifts to $^{19}$O$(\alpha,n)^{22}$Ne.  For a low $^{17}\mathrm{O}(n,\gamma)$ cross section, the transition sets in at a slightly lower value of $Y_e$ because the flow to $^{19}$O is inhibited. As a consequence, the exact $Y_e$ for the transition is also sensitive to  $^{17}\mathrm{O}(n,\gamma)$. The main contributors to the $RMSD$ in these regions are isotopes in the first $r$-process peak, including isotopes of Ge, As, and Se. 
The stable Se isotopes are made as neutron-rich Ni, Cu, and Zn isotopes. The impact thus reflects the change in the production of Fe-peak seed isotopes shortly after freeze-out from NSE.  

 However, in this region, the mass fractions of the first $r$-process peak elements barely reach $10^{-5}$ while the overall abundance pattern is dominated by the second $r$-process peak around $A=130$. 
In this region, we see a minor impact on $^{129}\rm{I}$, which changes by a few percent for different $^{17}\mathrm{O}(n,\gamma)$ reaction cross sections. This is interesting, because $^{129}\rm{I}$ is a radioactive isotope with a half-life of 15~Myr, signatures of which have been identified in early solar-system materials \citep{Cote_I129}, with an uncertainty at the percent level. 
 
\section{Conclusion} \label{sec:conclusion}
We have presented a new calculation for the direct neutron capture cross section $^{17}\mathrm{O}(n,\gamma)$ calculated with a microscopic approach that goes beyond previous, more phenomenological attempts. We find values of the cross section in the relevant energy range that are similar to a recent analysis but we confirm that it is significantly different from both the commonly used REACLIB library for astrophysical rates and the ENDF/B VIII.1 evaluated nuclear data library. We have analyzed the impact of the new reaction rate on weak $r$-process nucleosynthesis, important for the origin of the first $r$-process peak elements, including strontium ($Z=38$), which provide important observational signatures to identify the sites of the $r$ process. We find that the higher neutron capture cross section compared to the REACLIB library decreases the production of Sr and Zr, albeit our results show a less significant decrease than results by \citet{zhang2022}. We find significantly lower yields with our cross section than in with the cross section from ENDF/B VIII.1. 
We identify the shift in the reaction paths that leads to the impact of $^{17}\mathrm{O}(n,\gamma)$ in a relatively narrow astrophysical regime. The impact of $^{17}\mathrm{O}(n,\gamma)$ reaction cross section on the first $r$-process peak illustrates the importance of reviewing also the reaction on relatively light isotopes and updating commonly used reaction rate libraries as new experimental data and theoretical techniques become available.

\begin{acknowledgments}
NLA acknowledges support from the Vietnam National Foundation for Science and Technology Development (NAFOSTED) under grant number 103.04-2025.07. 
BML is supported by the U.S. Department of Energy, under Award Number DE-NA0004075.
The work of JSAP and AS was performed under the auspices of the U.S. Department of Energy by Lawrence Livermore
National Laboratory under Contract DE-AC52-07NA27344 with support from LDRD project 24-ERD-023. 
JSAP acknowledges support through the Cal-Bridge program and the Defense Science and Technology Internship program at LLNL.
This material is based in part upon work supported by the U.S. Department of Energy, Office of Science, Office of Nuclear Physics, under Work Proposal No. SCW0498.
Computing support for this work came in part from the LLNL
institutional Computing Grand Challenge program. We thank J. Escher and the whole RETRO team as well as, W.R. Hix, J.A. Harris and the ORNL summer interns for helpful discussions. 
\end{acknowledgments}

\bibliographystyle{apsrev4-2}
\bibliography{refs,astro_refs}

\end{document}